\documentclass[epj]{svjour}

\usepackage{graphicx}
\usepackage{fancyhdr}
\def\makeheadbox{{

 }}
\def\mytitle{Direct Detection of Cold Dark Matter } 
\def\myauthors{Laura Baudis}    
\def\mytype{Review}
\def\mysession{\myauthors}

\begin{document}
\title{Direct Detection of Cold Dark Matter}
\author{Laura Baudis 
\thanks{\emph{Email:} laura.baudis@physik.uzh.ch}%
 }            
\institute{Physik Institut, University of Zurich, Winterthurerstr. 190, CH-8057 Zurich}
%
\date{}
\abstract{
We know from cosmological and astrophysical observations that more than 80\% of the matter density 
in the Universe is non-luminous, or dark. This non-baryonic dark matter could be composed of neutral, 
heavy particles, which were non-relativistic, or 'cold', when they decoupled from ordinary matter. 
I will review the direct detection methods of these hypothetical particles via their interactions with 
nuclei in ultra-low background, deep underground experiments. The emphasis is on most recent results and 
on the status of near future projects.
} 

\maketitle
%

\section{Introduction}
\label{intro}

In the Standard Model of cosmology the Universe is flat and made of 4\% baryons, 20\% non-baryonic dark matter 
and 76\% dark energy \cite{turner}. Generic candidates for non-baryonic dark matter, which thus makes more than 80\% of the 
total matter density in the Universe, are axions \cite{raffelt} and 
weakly interacting massive particles (WIMPs) \cite{jkg}. WIMPs could have been thermally produced in the very early universe. 
If their mass and cross section is determined by the weak scale, the freeze-out relic density is around the observed 
value, $\Omega\sim$0.1 \cite{feng}. The best studied WIMP candidates are the neutralino, or the lightest supersymmetric particle,
and the lightest Kaluza-Klein particle, for instance the B$^{(1)}$, which is the first excitation of 
the hypercharge gauge boson in theories with universal extra dimensions. A recent comprehensive review of 
particle dark matter candidates has been provided by Bertone, Hooper and Silk \cite{gf}. A recent review on axions as 
cold dark matter candidates can be found in \cite{axions}. 
In this review I will focus on the direct detection of WIMPs, with emphasis on most recent experimental results and 
on the status of projects which are now under construction.

\section{Direct Detection of WIMPs}
\label{sec:1}

WIMPs can be detected directly, via their elastic collisions with nuclei in terrestrial targets \cite{goodman85}. 
The differential rate for WIMP elastic scattering off nuclei is given by  \cite{smith_lewin96}

\begin{equation}
\frac {dR}{dE_R}=N_{T}
\frac{\rho_{0}}{m_{W}}
                    \int_{v_{\rm min}}^{v_{\rm max}} \,d \vec{v}\,f(\vec v)\,v
                     \,\frac{d\sigma}{d E_R}\,  
\label{eq1}
\end{equation}

\noindent
where $N_T$ represents the number of the target nuclei,
$m_W$ is the WIMP mass and $\rho_0$
the local WIMP density in the galactic halo,
$\vec v$ and $f(\vec v)$ are the WIMP
velocity and velocity distribution function  in the Earth frame
 and ${d\sigma}/{d E_R}$ is the WIMP-nucleus differential cross section.
The nuclear recoil energy is 
$E_R={{m_{\rm r}^2}}v^2(1-\cos \theta)/{m_N}$,
where $\theta$ is the  scattering
angle in the WIMP-nucleus center-of-mass frame,
$m_N$ is the nuclear mass and $m_{\rm r}$ is the WIMP-nucleus
reduced mass. The velocity $v_{\rm min}$ is defined as 
$v_{\rm min} = (m_N E_{th}/2m_{\rm r}^2)^{\frac{1}{2}}$, where $E_{th}$
is the energy threshold of the detector, and 
$v_{\rm max}$ is the escape WIMP velocity in the Earth frame.  
The simplest galactic model assumes a Maxwell-Boltzmann distribution for the
WIMP velocity in the galactic rest frame, with a velocity dispersion 
of v$_{rms} \approx$ 270\,km\,s$^{-1}$ and an escape 
velocity of v$_{esc} \approx$ 650\,km\,s$^{-1}$. 

The differential WIMP-nucleus cross section has two 
separate components: an effective scalar coupling between the WIMP and the 
nucleus  and  an effective coupling between the spin of the WIMP and the total spin of the nucleus.
In general the coherent part dominates the interaction (depending however 
on the content of the neutralino) for target masses with A$\geq$30 \cite{jkg}.
The cross section  can be expressed as the sum of both contributions    
$\frac{d\sigma}{d E_R} \propto {\sigma_{SI}^0} F_{SI}^2(E_R) + {\sigma_{SD}^0} F_{SD}^2(E_R)$, where 
 $\sigma_{SI,SD}^0$ are the WIMP-nucleus 
cross sections in the limit of zero momentum transfer 
and $F_{SI,SD}^2(E_R)$ denote the nuclear form factors, 
expressed as a function of the recoil energy.

\begin{figure}[thbp]
\includegraphics[width=0.5\textwidth,angle=0]{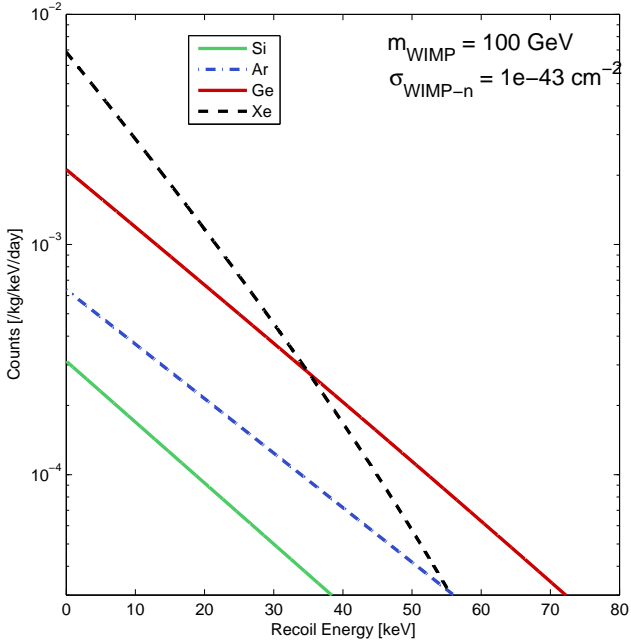}
\caption{Differential WIMP recoil spectrum for a WIMP mass of 100\,GeV 
and a WIMP-nucleon cross section $\sigma = 10^{-43}$ cm$^2$. The spectrum was 
calculated for illustrative nuclei such as Si (light solid), Ar (light dot-dashed), Ge (dark solid), Xe (dark dashed).} 
\label{fig1:rates}
\end{figure}

The left side of equation \ref{eq1} is the measured spectrum in a detector,
while the right side represents the theoretical prediction. 
It includes  WIMP properties which are completely unknown,
such as the mass $m_W$ and elastic cross section $\sigma_0$, 
quantities in principle accessible from astrophysics, such as the 
density of WIMPs in the halo $\rho_0$, the WIMP velocity distribution $f(v)$ and the escape
velocity (which are however still prone to large uncertainties) and
detector specific parameters: mass of target nucleus, energy
threshold and nuclear form factors.
The nuclear form factors become significant at large WIMP and nucleus 
masses, and lead to a suppression of the differential scattering rate. 
Figure \ref{fig1:rates} shows differential spectra for Si, Ar, Ge and Xe, 
calculated for a WIMP mass of 100\,GeV, a scalar WIMP-nucleon cross section of 
$\sigma = 10^{-43}$ cm$^2$, using standard halo parameters ( $\rho_0$ = 0.3 GeV/cm$^3$, 
$v_0$ = 220 km s$^{-1}$).

A WIMP with a typical mass between a few GeV and 1\,TeV 
will deposit a recoil energy below 50\,keV in a terrestrial detector.
The predicted event rates for neutralinos range from 10$^{-6}$ to 10
events per kilogram detector material and day \cite{jkg}.
Evidently, in order to observe a WIMP spectrum, 
low energy threshold, low background and high mass detectors are essential.
In such a detector, the recoil energy of the scattered nucleus is transformed into
a measurable signal, such as charge, light or phonons, 
and at least one of the above quantities is detected. Observing two  
signals simultaneously yields a powerful discrimination against background events,  
which are mostly interactions with electrons, as opposed to WIMPs and neutrons scattering 
off nuclei.
Even for experiments with good event by event discrimination, an absolute low background   
is still important, given the small size and rate of the expected WIMP signal. 
It can be achieved in both passive and active ways. Passive methods 
range from high material selection of detector
components to various specific shields against the natural radioactivity of the environment 
and against cosmic rays and secondary particles produced in their interactions.
Active background reduction implies an active shield, commonly a plastic or liquid 
scintillator, or a water Cerenkov shield surrounding the detector. An additional advantage 
is provided by a highly granular detector (or good timing and position resolution), 
since multiple scatters within the detector volume allow both background reduction 
and a direct measurement of the neutron background for experiments with an event-by-event 
discrimination against electron recoils. Finally, the position resolution leads  
to the identification of events clustered at the detector surfaces,  
which are quite unlikely to be WIMPs.

In order to convincingly detect a WIMP signal, a specific signature from a particle populating  
our galactic halo is important.
The Earth's motion through the galaxy induces both a seasonal variation of 
the total event rate \cite{fre86,fre88} and a forward-backward 
asymmetry in a directional signal \cite{spergel88,copi99}. 
The annual modulation of the WIMP signal arises because of the Earth's motion 
in the galactic frame, which is a superposition of the Earth's rotation
around the Sun and the Sun's rotation around the galactic center:

\begin{equation}
v_E = v_\odot + v_{\rm orb} \; \cos \gamma \; \cos \omega (t - t_0), 
\end{equation}

where $v_\odot = v_0$ + 12 km s$^{-1}$ ($v_0$ $\approx$ 220  km s$^{-1}$),  
$v_{\rm orb}$ $\approx$ 30 km s$^{-1}$ denotes 
the Earth's orbital speed around the Sun, the angle 
$\gamma \approx 60^0$ is the inclination of the Earth's orbital plane 
with respect to the galactic plane and $\omega = 2 \pi / 1 \mbox{yr}$, 
$t_0 =$ June  2$^{\rm nd}$.  

The expected time dependence of the count rate can  be approximated 
by a cosine function with a period of T = 1\,year and a phase of $t_0
=$ June  2$^{\rm nd}$:

\begin{equation}
S(t) = S_0 + S_m cos \omega (t - t_0),
\end{equation}

where $S_0$, $S_m$ are the constant and the modulated amplitude of the signal, 
respectively.
In reality, an additional contribution to S(t) from the
background must be considered.  The background should  be constant in time or at least 
not show the same time dependency as the predicted WIMP signal.
The expected seasonal modulation effect is very small 
(of the order of $v_{\rm orb}$/$v_0 \simeq $ 0.07), requiring
large masses and long counting times as well as an excellent long-term stability 
of the experiment.

A much stronger signature would be given by the ability to detect the axis and 
direction of the recoil nucleus.
It has been shown \cite{spergel88}, that the WIMP interaction rate as a function 
of recoil energy and angle $\theta$ between the WIMP velocity and
recoil direction (in the galactic frame) is:

\begin{equation}
\frac{d^2R}{dE_R d cos\theta} \propto \rm{exp}\left[-\frac{(v_\odot cos\theta - v_{min})^2}{v_0^2} \right],  
\end{equation}

where v$_{min}^2$ = (m$_N$ + m$_W$)$^2$E$_R$/2m$_N$m$_W^2$ 
and v$_0^2$ = 3 v$_\odot^2$/2.
The forward-backward asymmetry yields a large effect of the order of 
$\mathcal{O}$(v$_\odot$/v$_0$)$\approx$ 1. Thus fewer events are needed 
to discover a WIMP signal than in the case where the seasonal modulation 
effect is exploited \cite{copi99}.
The challenge is to build massive detectors capable of detecting the direction of the 
incoming WIMP. At present, only one such experiment exists: two DRIFT-II modules \cite{drift}, with 
an active mass of $\sim$167\,g  of CS$_2$ each (1 m$^3$ target), with larger modules 
being under consideration.

\section{Experiments}

\begin{figure}[thbp]
\includegraphics[width=0.5\textwidth,angle=0]{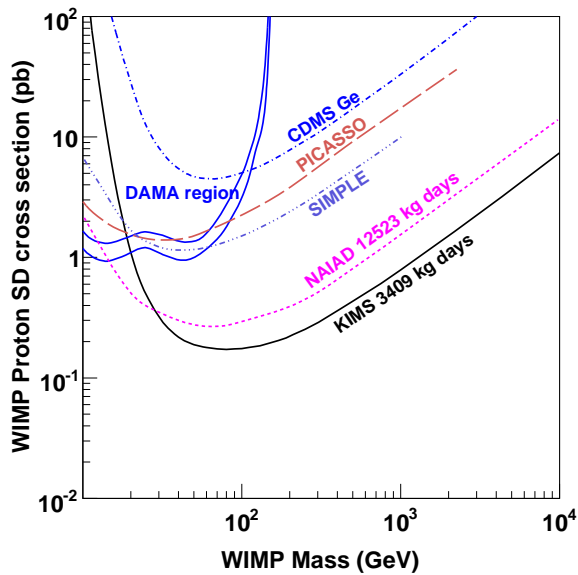}
\caption{Experimental results for spin-dependent WIMP nucleon cross sections 
versus WIMP mass, for pure proton couplings  (Figure from \cite{kims}).} 
\label{fig2:limitsSD}

\end{figure}

A large range of techniques and target materials are currently being used to directly 
search for WIMPs. Since this review is focused on most recent results, 
I refer to \cite{gabriel05,rick04} for more exhaustive overviews. 
Cryogenic experiments operated at sub-Kelvin temperatures (CDMS \cite{cdms}, 
CRESST \cite{cresst}, EDELWEISS \cite{edelweiss}) and  detectors based on liquid noble elements 
such as Xe (XENON10 \cite{xe10-prl}, ZEPLIN-II \cite{zeplin2}) and Ar (WARP \cite{warp})
are currently leading the field with sensitivities to 
WIMP-nucleon cross sections (for spin independent couplings) ranging from 1.6$\times$10$^{-43}$cm$^2$ 
(CDMS ~\cite{cdms-prl}) to 8.8$\times$10$^{-44}$cm$^2$ (XENON10 \cite{xe10-prl}) at a WIMP mass of 100 GeV/c$^2$. 
Projects which are already under operation, or will be so in the near future, 
such as ZEPLIN-III \cite{zeplin3}, XENON100 \cite{xenon100}, ArDM \cite{ardm}, and the upgraded versions 
of CDMS, CRESST, EDELWEISS and WARP should improve these results by at least one order of magnitude.
Recent progress with solid CsI scintillator detectors, by the Korea Invisible Mass Search (KIMS) \cite {kims}, 
allowed to set the most stringent limits for the spin-dependent, pure proton couplings (Figure 
\ref{fig2:limitsSD}). The total exposure with 4 CsI crystals (34.8 kg) of 3409 kg days 
allowed to probe the spin-dependent cross section down to 
1.8$\times$10$^{-37}$cm$^2$ at a WIMP mass of 100 GeV/c$^2$, and thus to test the DAMA \cite{dama} signal region 
with crystal scintillators containing iodine for the first time \cite{kims}.
LIBRA, a 250 kg NaI experiment operated by the DAMA collaboration at LNGS since 2003 \cite{libra}, will 
likely present first results in 2008. With improved statistics, higher light yields and 
lower backgrounds, they will either confirm, or rule out the annual modulation of 
events previously observed by the DAMA NaI crystals \cite{dama}. An independent cross-check 
with NaI will be provided by ANAIS \cite{anais} with more than 100 kg of NaI detectors.
Other techniques, such as superheated liquids are rapidly evolving and may be scalable 
to large, inexpensive room-temperature detectors with sensitivities to both scalar and axial-vector 
couplings (COUPP \cite{coupp}, PICASSO \cite{picasso} and SIMPLE \cite{simple}).

\subsection{Cryogenic Detectors at mK Temperatures}

Cryogenic calorimeters are meeting crucial characteristics of a successful WIMP 
detector: low energy threshold ($<$10\,keV), excellent energy resolution ($<$1\% at 10\,keV)  
and the ability to differentiate nuclear from electron recoils on an event-by-event basis. 
Their development was driven by the exciting possibility of doing a calorimetric energy 
measurement down to very low energies with unsurpassed energy resolution. 
Because of the T$^3$ dependence of the heat capacity of a dielectric crystal, 
at low temperatures a small energy deposition can significantly change the temperature 
of the absorber. The change in temperature is measured either after the phonons 
(or lattice vibration quanta) reach equilibrium, or thermalize, or when they are still 
out of equilibrium, or athermal, the latter providing additional information 
about the location of an event.

\begin{figure}[thbp]
\includegraphics[width=0.45\textwidth,angle=0]{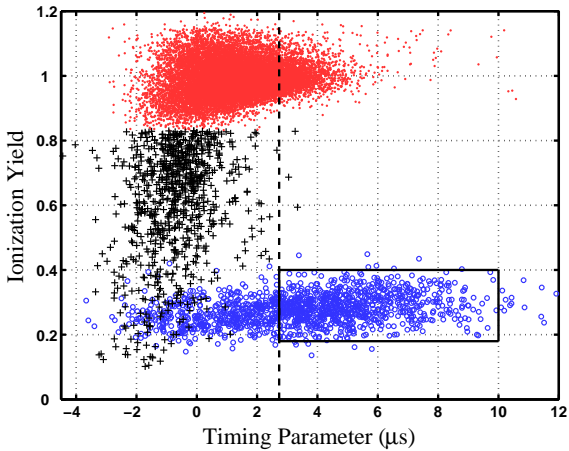}
\caption{Ionization yield versus phonon timing parameter for $^{133}$Ba gamma calibration events (dots and crosses)
and $^{252}$Cf neutron calibration events (circles). Low-yield $^{133}$Ba events (crosses) 
have small values of the timing parameter, and the dashed vertical line indicates a timing 
cut, resulting in a high rate of nuclear recoil efficiency and a low rate of misidentified surface events (Figure from \cite{cdms-prl}).} 
\label{fig3:cdms_rt}
\end{figure}

\noindent
{\bf CDMS and SuperCDMS:} the Cold Dark Matter Search experiment operates 30 low-temperature Ge and Si detectors at the 
Soudan Underground Laboratory in Minnesota (at a depth of 2080 m.w.e.).  The high-purity 
Ge and Si crystals are 1\,cm thick and 7.6\,cm in diameter, and have a mass of 250\,g and 
100\,g, respectively. Superconducting 
transition edge sensors photolitographically patterned onto one of the crystal surfaces 
detect the athermal phonons from particle interactions. The phonon sensors are divided into 
4 different channels, allowing to reconstruct the x-y position of an event with a resolution 
of $\sim$1\,mm. If an event occurs close to the detector's surface, the phonon signal is faster 
than for events far from the surface, because of phonon interactions in the thin metallic films.
The risetime of the phonon pulses, as well as the time difference between the charge and phonon 
signals allow to reject surface events caused by electron recoils. Figure~\ref{fig3:cdms_rt} 
shows the ionization yield (ratio of ionization to recoil energy) versus the sum of above timing 
parameters  for electron recoil events (collected with a $^{133}$Ba source) and nuclear recoil events 
(collected with a $^{252}$Cf source). Events below a yield around 0.75 typically occur within 
0-30\,$\mu$m of the surface, and can be effectively discriminated  while preserving a large part of 
the nuclear recoil signal.

Charge electrodes are used for the ionization measurement. They are divided into an inner disk, 
covering 85\% of the surface, and an outer ring, which is used to reject events near the 
edges of the crystal, where background interactions are more likely to occur. 
The discrimination against the electron recoil background is based on the fact that 
nuclear recoils (caused by WIMPs or neutrons) produce fewer charge pairs than electron 
recoils of the same energy. The ionization yield is about 0.3 in Ge, and 0.25 in Si for 
recoil energies above 20\,keV. Electron 
recoils with complete charge collection show an ionization yield of $\approx$1.  
For recoil energies above 10\,keV, bulk electron recoils are rejected with $>$99.9\% efficiency,
and surface events are rejected with $>$95\% efficiency.
The two different materials are used to distinguish between WIMP and neutron interactions by 
comparing the rate and the spectrum shape of nuclear recoil events.

\begin{figure}[thbp]
\includegraphics[width=0.5\textwidth,angle=0]{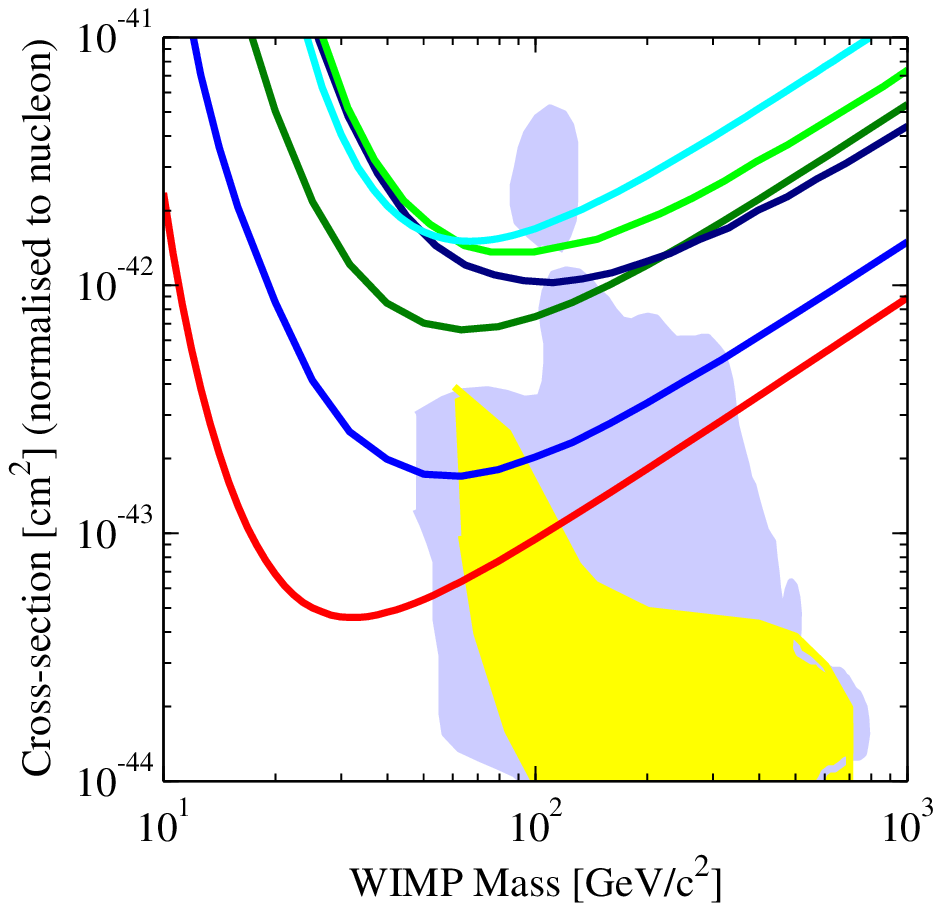}
\includegraphics[width=0.5\textwidth,angle=0]{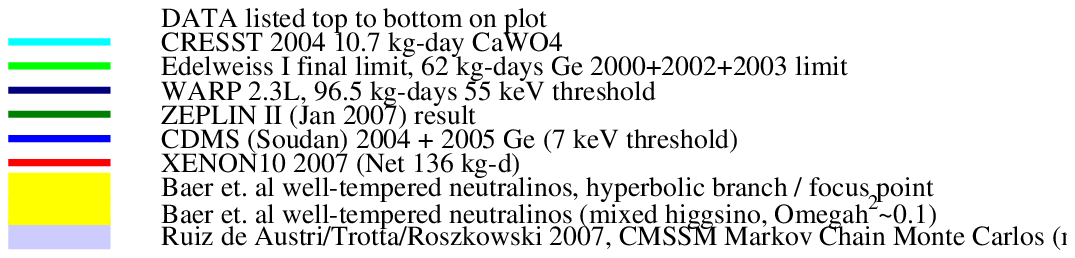}
\caption{Experimental results and theoretical predictions for spin-independent WIMP nucleon cross sections 
versus WIMP mass (Figure generated with limitplotter \cite{limitplotter}).} 
\label{fig3a:limitsSI}
\end{figure}

A stack of six Ge or Si detectors together with the corresponding cold electronics is named a 'tower'.
Five towers are currently installed in the 'cold volume' at Soudan, shielded by about 3\,mm of Cu, 
22.5\,cm of Pb, 50\,cm of polyethylene and by a 5\,cm thick plastic scintillator detector  
which identifies interactions caused by cosmic rays penetrating the Soudan rock.
In 2004, two towers  were operated for 74.5 live days at Soudan, yielding an exposure 
of 34\,kg\,d in Ge and 12\,kg\,d in Si in the 10-100\,keV nuclear recoil energy range. 
One candidate nuclear recoil event at 10.5 keV was observed in Ge, while no 
events were seen in the Si data \cite{cdms-prl}. This result was consistent with the expected background 
from surface events, and resulted in a new upper limit on spin-independent WIMP-nucleon 
cross sections in Ge of 1.6$\times$10$^{-43}$cm$^2$ at the 90\%CL at a WIMP mass of 60\,GeV/c$^2$ 
(Figure \ref{fig3a:limitsSI}). 

The accumulated statistics of the two completed 5-tower runs at Soudan (the first one from October 21, 06 to March 21, 07,  
the second one from April 20 to July 16, 2007) amounts to about 660\,kg\,day in Ge; the third 
run has been started on July 21st, 2007 and is currently ongoing (the exposure is approaching 
1\,ton\,day in Ge).  The analysis of the first two data sets is already in progress, with 
an expected sensitivity around 4$\times$10$^{-44}$cm$^2$ at a WIMP mass of 100 GeV/c$^2$. 

\begin{figure}[thbp]
\includegraphics[width=0.45\textwidth,angle=0]{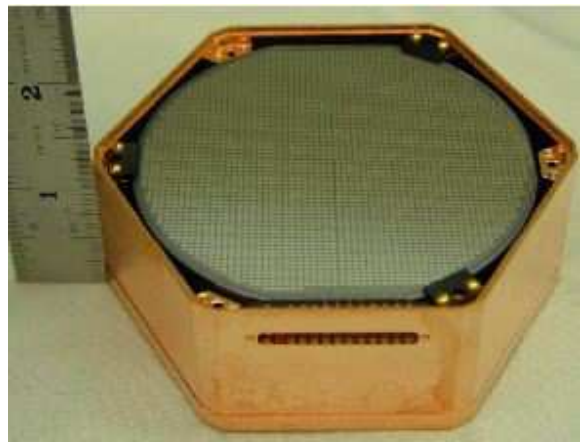}
\caption{A SuperCDMS detector, 7.5 cm in diameter, and 2.5 cm thick. The WIMP target mass is 640\,g for a Ge, 
and 250\,g for a Si crystal. The phonon collection area was increased by a factor of two relative to CDMS ZIPs, 
and new H-a-Si electrodes suppress the charge back diffusion.} 
\label{fig4:superZIP}
\end{figure}

The SuperCDMS-25 experiment \cite{brink05}, consisting of 42 new, 640 g Ge ZIP detectors, will operate at SNOLab, as 
a first step towards the realization of a ton-scale cryogenic experiment (Figure \ref{fig4:superZIP} shows a 
picture of a new ZIP detector). The goal of SuperCDMS-25 is to improve 
existing best limits by more than two orders of magnitude, with improved background rejection, lower absolute background 
and advanced timing analysis techniques for the new 1'' thick ZIP detectors.

\noindent
{\bf EDELWEISS:} the EDELWEISS experiment operates Ge bolometers at 17\,mK in the Laboratoire 
Souterrain de Modane, at  4800\,m.w.e. The detectors are further shielded by 30\,cm 
of paraffin, 15\,cm of Pb and 10\,cm of Cu. They simultaneously detect 
the phonon and the ionization signals, allowing a discrimination against bulk electron 
recoils of better than 99.9\% above 15\,keV recoil energy. The charge signal is measured by Al 
electrodes sputtered on each side of the crystals, the phonon signal by a neutron transmutation 
doped (NTD) heat sensor glued onto one of the charge collection electrodes. The NTD sensors read out 
the thermal phonon signal on a time scale of about 100\,ms. 
Between 2000-2003, EDELWEISS performed four physics runs with five 320\,g Ge crystals, accumulating 
a total exposure of 62\,kg\,days \cite{edelweiss}. Above an analysis threshold of 20\,keV, 
a total of 23 events compatible with nuclear recoils have been observed. 
The derived upper limits on WIMP-nucleon couplings under the hypothesis that all above events are caused
by WIMP interactions, and for a standard isothermal halo, are shown in Figure~\ref{fig3a:limitsSI}.
The EDELWEISS-I experiment has ceased running in March 2004, in order to allow the upgrade to a second phase, 
with an aimed sensitivity of 10$^{-44}$cm$^2$. The new 50 liter low-radioactivity cryostat is able to house 
up to 120 detectors. Because of the inability of slow thermal detectors (Ge/NTD)  
to distinguish between low-yield surface events and nuclear 
recoils and the inherent radioactivity of NTD sensors, the collaboration has developed a new design 
based on NbSi thin-film sensors (Ge/NbSi). These films, besides providing a lower mass and radioactivity per sensor, 
show a strong difference in the pulse shape, depending on the interaction depth of an event \cite{nader01}.
Four runs with 8 detectors had been performed in 2006, and have provided various system checks; 
in 2007, 21 Ge/NTD (330 g each) and 4 Ge/NbSi (400 g each) detectors have been mounted, 
and first neutron and gamma calibration data, 
as well as a low-background data set, have been taken. Before discrimination for nuclear recoils, the 
preliminary background is about a factor of two lower than the one of EDELWEISS-I detectors. 
The electron versus nuclear recoil discrimination of these new detectors is similar to the one obtained with 
EDELWEISS-I. The goal of this phase is to reach a sensitivity of 
 10$^{-43}$cm$^2$ by July 2008, and  10$^{-44}$cm$^2$ by 2009/2010 \cite{edelw-2}.

\noindent
{\bf CRESST:} The CRESST collaboration has developed cryogenic detectors based on CaWO$_4$ crystals, 
which show a higher light yield at low temperatures compared to other scintillating materials.  
The detectors are also equipped with a separate, cryogenic light detector made of a 30$\times$30$\times$0.4\,mm$^3$
silicon wafer, which is mounted close to a flat surface of the CaWO$_4$ crystal. 
The temperature rise in both CaWO$_4$ and light detector is measured with  tungsten superconducting 
phase transition thermometers, kept around 10\,mK, in the middle of their transition between the superconducting 
and normal conducting state. 
A nuclear recoil in the 300\,g CaWO$_4$ detector 
has a different scintillation light yield than an electron recoil of the same energy, allowing to 
discriminate between the two type of events when both the phonon and the light signals are observed. 
The advantage of CaWO$_4$ detectors is their low energy threshold in the phonon signal, and the fact 
that no light yield degradation for surface events has been detected so far. 
However, about 1\% or less of the energy deposited in the CaWO$_4$ is seen as scintillation 
light \cite{cresst}. Only a few tens of photons are emitted per keV electron recoil, 
a number which is further 
diminished for nuclear recoils, because of the involved quenching factor. 
The quenching factor of oxygen nuclear recoils for scintillation light is around 
13.5\% relative to electron recoils \cite{cresst}, leading to a rather high effective recoil energy threshold for  
the detection of the light signal. While neutrons will scatter predominantly on oxygen nuclei, 
it is expected that WIMPs will more likely scatter on the heavier calcium and tungsten. 
The most recent CRESST results \cite{cresst} were obtained by operating two 300\,g CaWO$_4$ detectors at the Gran Sasso 
Underground Laboratory (3800 m.w.e) for two months at the beginning of 2004. 
The total exposure after cuts  was 20.5\,kg\,days. 
The energy resolution in the phonon channel was 
 1.0\,keV (FWHM) at 46.5\,keV, the low-energy gamma line being provided by an external $^{210}$Pb contamination.  
A total of 16 events were observed in the 12\,keV - 40\,keV recoil energy region, 
a number which seems consistent with the expected neutron background, since 
the experiment had no neutron shield at that time. 
No phonon-only events (as expected for WIMP recoils on tungsten) were observed between 12\,keV - 40\,keV 
in the module with better resolution in the light channel, yielding a limit on coherent WIMP 
interaction cross sections similar to the one obtained by EDELWEISS.
CRESST had stopped taking data in March 2004, and the experiment was upgraded with a 66-channels SQUID read-out system
(for 33 detectors, including new wiring), with a neutron moderator made out of polyethylene, 
and with a scintillator muon veto. A commissioning run with 10 detector modules (3 kg) has provided  
a test of the new system, as well as first low-background data. Results on WIMP-nucleon 
cross sections will be presented after the in situ neutron calibration. Preliminary limits 
are around 10$^{-42}$cm$^2$ at a 100 GeV/c$2$ WIMP mass \cite{seidel07}.
The final goal of this phase is to operate 33 CaWO$_4$ detector modules (10 kg), reaching 
a sensitivity of 10$^{-44}$cm$^2$.
\noindent
{\bf EURECA:}  the European Underground Rare Event search with Calorimeter Array is a design study to develop a 100\,kg-1\,ton 
cryogenic, multi-target experiment in Europe, based on  CRESST and EDELWEISS expertise \cite{gabriel05}.

\begin{figure}[thbp]
\includegraphics[width=0.45\textwidth,angle=0]{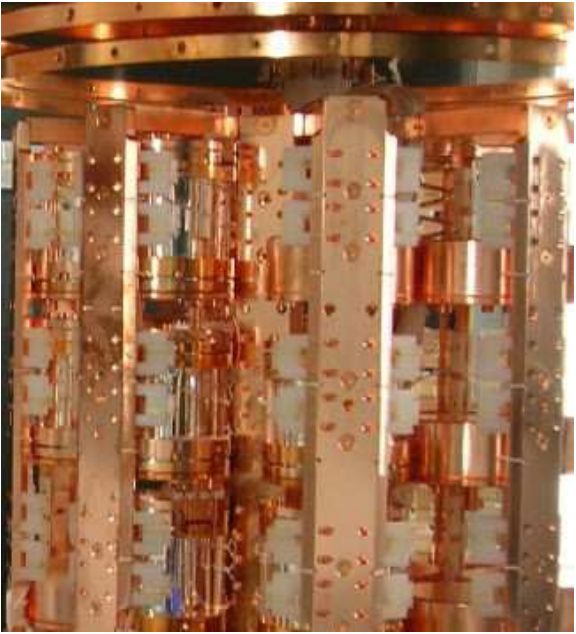}
\caption{The CRESST 3\,kg detector array (10 modules), currently under operation at LNGS. The detector support structure 
is made of special low-background, low heat leak copper (Figure from \cite{seidel07}).} 
\label{fig4a:cresst}
\end{figure}

\subsection{Liquid Noble Element Detectors}
\label{liquid_noble}

Liquid noble elements have excellent properties as dark matter targets and are now being used to build large, 
homogeneous and position sensitive WIMP  detectors. The main materials are neon (CLEAN \cite{clean}), 
argon (ArDM \cite{ardm}, WARP \cite{warp}, DEAP \cite{deap}) and xenon  (XENON10 \cite{xenon10}, XENON100 \cite{xenon100}, 
ZEPLIN-II \cite{zeplin2}, ZEPLIN-III \cite{zeplin3}, LUX \cite{lux}, XMASS \cite{xmass}).
These are intrinsic scintillators, having high scintillation ($\lambda$ = 85 nm (Ne), 128 nm (Ar), 175 nm (Xe)) 
and ionization yields because of their low ionization potentials (21.56~eV in Ne, 15.75~eV in Ar and 12.13~eV in Xe) 
\cite{elena-book}.
Scintillation is produced by the formation of excimer states, which are 
bound states of ion-atom systems. If a high electric field ($\sim$1 kV/cm) is applied, ionization electrons can also
be detected, either directly or through the secondary process of proportional scintillation.
The elastic scattering of a WIMP produces a low-energy nuclear recoil,  which loses its energy through ionization and scintillation 
(and of course heat, which is not detected in this case). 
Both signals are suppressed when compared to an electron recoil of the same energy, but by different amounts, allowing 
to use the ratio for distinguishing between electron and nuclear recoils. The scintillation and ionization yields
depend on the drift field and on the energy of the recoil and must be known down to low nuclear recoil energies. 
Because of the short wavelength of their scintillation light, the operation of neon and argon as scintillators requires wavelength shifters. 
Both materials contain only even isotopes, and thus don't provide sensitivity to axial-vector WIMP-nuclei interactions. 
An advantage of liquid noble elements is that they can easily be purified. 
While neon is free of any radioactive isotopes, natural argon contains about 1\,Bq/kg of $^{39}$Ar, which is a beta emitter 
with an endpoint of 565\,keV and a half-life of 269 years (yielding an integrated event rate 
of 1\,kHz in a 1 ton detector). Argon with low $^{39}$Ar content, or isotopic separation, as well an an 
excellent background rejection ($\sim$10$^8$) is therefore required for operating large liquid argon detectors.
Natural xenon does not contain any long-lived radioactive isotopes (apart from the double beta emitter $^{136}$Xe which  
has a half-life larger than 0.5$\times$10$^{21}$years). However the trace content of $^{85}$Kr (which is a beta emitter with  
an endpoint of 687\,keV and a half-life of 10.76 years) needs to be separated to few 100\,ppt levels (for a 100\,kg dark matter 
detector). Such levels can be achieved by distillation (as demonstrated by the XMASS collaboration \cite{xmass}) 
or by the adsorption process on a charcoal column \cite{elena-book}. 
The high mass of the Xe nucleus is favorable for WIMP scalar interactions (provided that a low energy threshold can be achieved) 
and the presence of two isotopes with unpaired neutrons ($^{129}$Xe, spin 1/2, 26.4\% and $^{131}$Xe, spin 3/2, 21.2\%) 
offers excellent sensitivity to the WIMP spin-dependent interaction. The high density (3\,g/cm$^3$) and high atomic number (Z=54, A=131.3), 
allow to build self-shielding, compact dark matter detectors.

\begin{figure}[thbp]
\includegraphics[width=0.5\textwidth,angle=0]{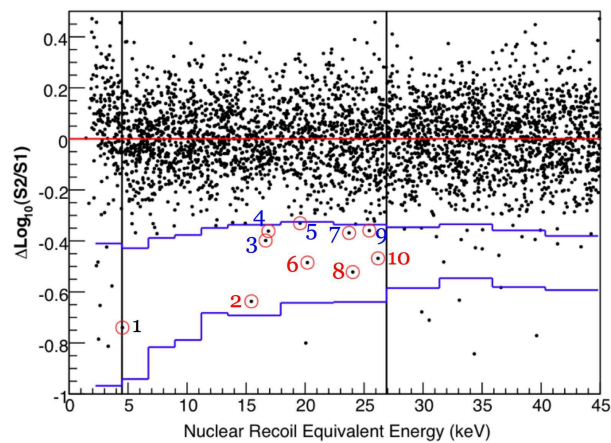}
\caption{XENON10 results from 58.6 live-days of WIMP-search in a fiducial mass of 5.4\,kg of LXe. The WIMP 
search region was defined between the vertical, and horizontal lines, with ten events 
observed in this region (Figure from \cite{xe10-prl}).} 
\label{fig6:xenon}
\end{figure}

\noindent
{\bf XENON10:} the XENON10 collaboration has operated a 15\,kg (active mass) dual phase detector time projection chamber in the Gran Sasso Underground 
Laboratory (LNGS), in WIMP search mode from August 24, 2006 to February 14, 2007. 
It uses two arrays of UV-sensitive photomultipliers (PMTs) to detect the prompt and proportional light signals induced 
by particles interacting in the sensitive xenon volume. 
The bottom array of 41 PMTs is located below the cathode, fully immersed in LXe, and mainly detects the 
prompt light signal. The 48 PMTs of the top array are located in the cold gas above the liquid, detecting the proportional 
light. XENON10 has full 3D position sensitivity:
the time separation between the two pulses of direct and proportional light (with a of maximum 75 $\mu$s)  provides the event depth of 
an interaction ($<$1\,mm resolution), 
the hit pattern in the top PMT array providing the x-y position (few mm resolution).
The position sensitivity, along with the self shielding of liquid xenon,  serve, along with the prompt versus 
proportional light ratio, as an important background rejection feature. 
The gamma background of XENON10 is at the level of 0.6~events/kg/d/keV after fiducial volume cuts (the fiducial mass 
is 5.4\,kg).
The energy threshold was 4.5 keV nuclear recoil energy (keVr) with a light yield of 0.7 photoelectrons/keVr. 
XENON10 was calibrated with both gamma and and AmBe neutron source, and stable operation had been demonstrated over many months 
\cite{xenon10}. Based on calibration, and on a period of  non-blind WIMP search data, 
the WIMP search region was defined between 4.5 and 29.6 keV nuclear recoil energy, 3$\sigma$ below the mean of 
the nuclear-recoil band (thus at 50\% nuclear recoil acceptance, see Figure \ref{fig6:xenon}). 
From a total of 1800 events in the 58.6 live-days of blind WIMP search data, 10 events were observed in the 
WIMP search region, with 7.0$^{+1.4}_{-1.0}$ events expected based on statistical (Gaussian) leagake alone. 
Given the uncertainty in the number of estimated leagake events from electron recoils (no neutron induced recoil 
events were expected for above exposure), conservative limits with no background subtraction were calculated 
for spin-independent WIMP cross sections. The 90\% C.L. upper limit at a WIMP mass of 100 GeV/c$^2$  is 
8.8$\times$10$^{-44}$cm$^2$ \cite{xe10-prl}.

\begin{figure}[thbp]
\includegraphics[width=0.5\textwidth,angle=0]{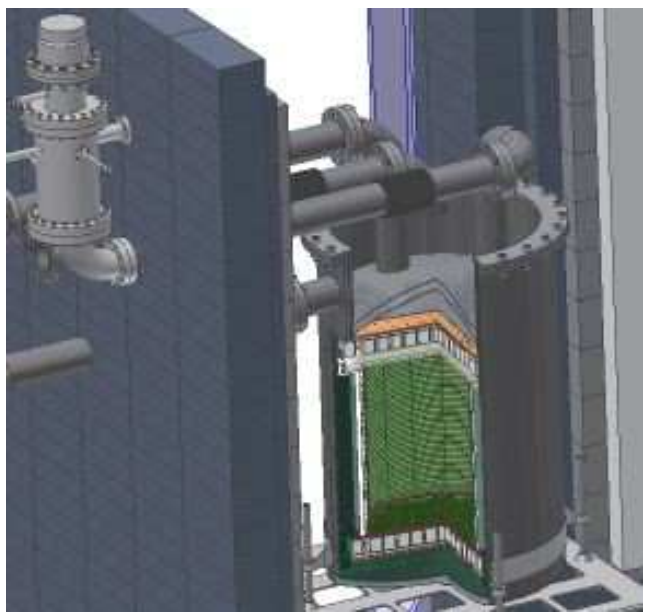}
\caption{Schematic view of the XENON100 detector in its low-background shield at LNGS. The high-voltage and signal feed-throughs, 
as well as the cryocooler are moved outside the Pb/PE shield.} 
\label{fig6a:xenon100}
\end{figure}

\noindent
{\bf XENON100 and LUX:} the XENON100 collaboration plans to install a 150 kg (70 kg WIMP target) detector at LNGS in early 2008.
The principle is similar to XENON10, with a few important differences: an active, LXe veto shield will surround the inner WIMP target, 
the signal and high-voltage feed-throughs, as well as the cryocooler are moved outside the Pb/PE shield, the double-walled 
stainless steel cryostat is made of low-radioactivity steel with $<$2 mBq/kg $^{238}$U and $^{232}$Th content. 
A total number of 250 low-radioactivity 1'' square PMTs ($\sim$0.2 mBq/PMT for $^{238}$U and $^{232}$Th) will view the WIMP target and the LXe shield.
With an increased mass, and a strongly reduced background, XENON100 is expected to reach a sensitivity of 
$\leq$9$\times$10$^{-45}$cm$^2$ in 2008.  Figure \ref{fig6a:xenon100} shows a schematic view of the XENON100 detectors in its 
low-background shield.
In the US, a 300 kg (100 kg fiducial mass) dual-phase Xe detector (LUX \cite{lux}), to be operated in a 
6\,m diameter water shield at DUSEL \cite{dusel}, has been proposed. A prototype detector is currently under construction at CWRU.

\begin{figure}[thbp]
\includegraphics[width=0.45\textwidth,angle=0]{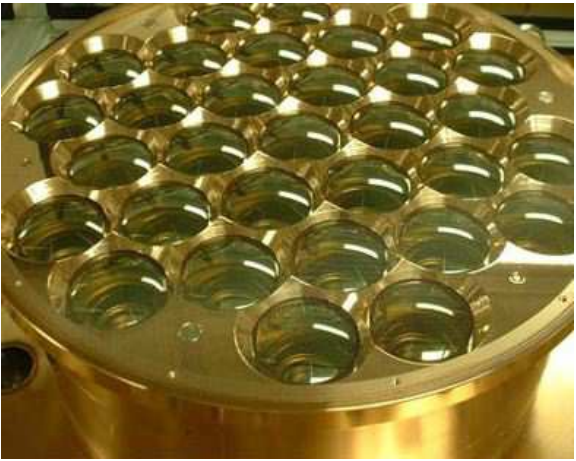}
\caption{The array of 31 2'' PMTs of the ZEPLIN-III experiment, which is currently under 
operation at the Boulby Lab in the UK (Figure from \cite{zeplin3a}).} 
\label{fig7:zep3}
\end{figure}

\noindent
{\bf ZEPLIN-II and ZEPLIN-III:} ZEPLIN-II is a two-phase LXe experiment, which is operated at the Boulby Underground 
Laboratory in the UK. The principle is similar to XENON10, with a LXe mass of 31\,kg 
viewed by one  PMT array of seven 13\,cm diameter PMTs. The depth of liquid xenon is 14\,cm, corresponding 
to a maximum drift time of 73\,$\mu$s for electrons at a drift field of 1 kV/cm. 
The experiment was calibrated with both gamma and neutron sources, which have helped to define 
the WIMP search region at 50\% nuclear recoil acceptance. The first data run had a total live time of 31.2 days, 
the fiducial mass after radial cuts to reject events coming from the PTFE walls was 7.2 kg. In the energy region 5-20 keV electron 
equivalent, 29 events were observed in the nuclear recoil acceptance region, with an expected background from gamma-ray 
and radon progeny events of 28.6$\pm$4.3 events. An upper limit of 10.4 nuclear recoil events at the 90\% C.L. was used
to derive a WIMP-nucleon spin-independent cross section limit with a minimum of 6.6$\times$10$^{-43}$cm$^2$ at 
a WIMP mass of 60 GeV/c$^2$  \cite{zeplin2}. A second run with a reduced radon   component is currently ongoing, the expected 
sensitivity is 1$\times$10$^{-43}$cm$^2$  for 5 months of stable underground operation.
The ZEPLIN-III detector is operating with about 8 kg of active liquid xenon target, viewed by 31 2'' PMTs \cite{zeplin3}, in its low-background 
shield at the Boulby Lab since June 2007. First results are expected in 2008, with a sensitivity close to 1$\times$10$^{-43}$cm$^2$
at a WIMP mass of 100 GeV/c$^2$. Figure \ref{fig7:zep3} shows the ZEPLIN-III PMT array covered by the electrode grid.

\noindent
{\bf ELIXIR} is a design study for a ton liquid xenon experiment in Europe, based on the technology, and experience gained with 
XENON and ZEPLIN.

\noindent
{\bf WARP:}
The WARP collaboration has operated a 2.3\,l two-phase argon drift chamber at LNGS in 2006. The liquid and gaseous argon 
(commercial grade for this first phase) volumes are viewed by an array of seven 2'' PMTs placed in the gas above the liquid, 
and coated with Tetra-Phenyl-Butadiene (TPB) which 
shifts the VUV photons emitted by the scintillating Ar to the PMT sensitive spectrum. The chamber was operated at a 
drift field of 1 kV/cm, for a total drift length of 7.5 cm. The trigger for an event was provided by the coincidence of at least 3 PMTs, 
with $\geq$1.5 p.e. in each PMT, with an efficiency close to unity above 20 p.e. (corresponding to 16 keV nuclear recoil energy). 
The live-time of the WIMP search run amounted to 52.8 days, corresponding to an exposure of 96.5 kg days. The fiducial volume 
was chosen to be 1.32 l (1.83 kg). The WIMP signal regions has been determined by an in situ neutron calibration with an AmBe source. 
Two different energy windows were selected for the WIMP search analysis: 40-60 keVr and 60-130 keVr; while 8 events are observed 
in the first window (of which 5 are single-hit events; these events could either be neutron, 
or electron-recoils leaking into the WIMP search region), none is  observed in the second energy region. The 90\% C.L. limit on WIMP-nucleon cross 
sections has been derived from the observation of zero events above 55 keVr, at the level of 1.2$\times$10$^{-42}$cm$^2$ at 
a WIMP mass of 100 GeV/c$^2$  \cite{warp}. By using both the pulse shape of each event, and 
the ratio of proportional to primary scintillation light, WARP has reached a background discrimination power of 3$\times$10$^{-7}$ 
above a threshold of 50 p.e. at 50\% nuclear recoil acceptance \cite{warp}. Since the background of the initial run was dominated by $^{39}$Ar decays, 
the collaboration is operating the chamber with isotopically separated argon \cite{warp-39ar}.
At the same time, a much larger detector with 140 kg of target material is under construction at LNGS.

\begin{figure}[thbp]
\includegraphics[width=0.45\textwidth,angle=0]{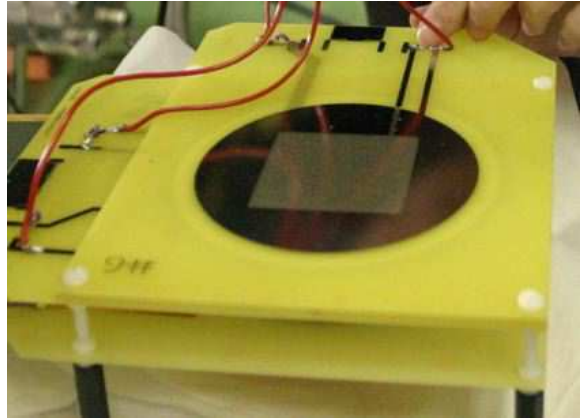}
\caption{A prototype array for a two-stage large electron multiplier for the ArDM experiment. The final 
LEM charge readout system will be segmented, allowing to determine the x-y position of events (Figure from \cite{ardma}).} 
\label{fig7a:ardm}
\end{figure}

\noindent
{\bf ArDM:}
ArDM is a ton-scale (850\,kg target mass) dual-phase argon detector, currently under construction at CERN. 
The scintillation light produced by a particle interacting in the sensitive volume is detected by an array of 14 TPB coated, 
20.2\,cm diameter low-background photomultiplier tubes,  which are located below the cathode grid in the liquid phase \cite{ardm}. 
The electrons are drifted over a drift length of about 120\,cm to the liquid-gas interface with the aid of an electric field, 
which is kept homogeneous with a series of field shaper rings, to which a Greinacher circuit provides the correct potentials. 
Once in the gas phase, the electrons are multiplied with a two-stage large electron multiplier (LEM), allowing to determine the 
x-y position of each event (the z position is inferred from the electron drift time). 
After the testing the full detector system at CERN in 2007, and determination of the discrimination power against 
electron recoils, an underground deployment, possibly at the Canfranc Lab, is planned. The goal is to test WIMP-nucleon 
cross sections in the range 1$\times$10$^{-42}$cm$^2$--1$\times$10$^{-44}$cm$^2$, depending on the exposure and 
the achieved background levels \cite{ardm}. Figure \ref{fig7a:ardm} shows a prototype for a 2-stage ArDM  LEM.

\subsection{Superheated Liquid Detectors}
An approach to build large WIMP detectors is to use superheated liquids. An energy deposition can destroy the 
metastable state, leading to the formation of bubbles, which can be detected and recorded both acoustically and optically. 
Since a minimal energy deposition is required to induce a phase-transition, these detectors are so-called threshold devices. 
The operating temperatures and pressure can be adjusted such that only nuclear recoils (large stopping powers $dE/dx$) 
lead to the formation of bubbles, making them insensitive to electron recoils coming from gamma interactions.  
Alpha-decays from radon and its progenies are still a problem, the current generation of experiments is attempting 
to minimize these to negligible levels. 
Two experiments (SIMPLE \cite{simple} and PICASSO \cite{picasso}) use superheated liquid droplets (C$_2$ClF$_5$ and C$_4$F$_{10}$) 
suspended in a water based gel. COUPP \cite{coupp} uses a large CF$_3$I liquid target 
operated in bubble chamber mode. 

\begin{figure}[thbp]
\includegraphics[width=0.5\textwidth,angle=0]{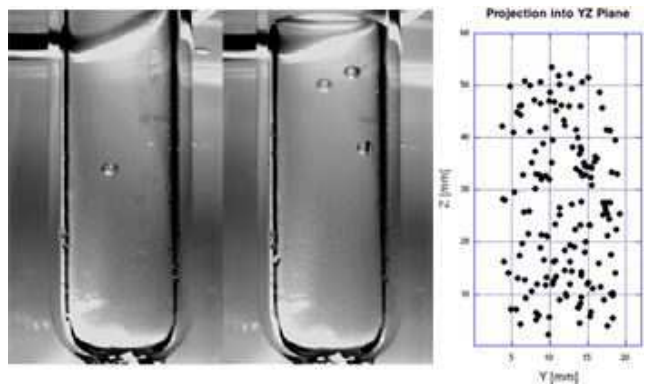}
\caption{The COUPP 2\,kg prototype. Left: a single bubble, as expected for WIMP scatters; middle: multiple bubbles 
as expected from neutron recoils; right: projection of events in the y-z-plan after 3D position reconstruction 
(Figure from \cite{couppa}).} 
\label{fig8:coupp}
\end{figure}

\noindent
{\bf COUPP: } the COUPP collaboration has installed and operated a  2\,kg CF$_3$I prototype chamber in the NuMI tunnel at Fermilab (300 m.w.e.) 
starting in 2005. 
The presence of fluorine and iodine in the target material allows sensitivity to both axial-vector and coherent WIMP-nucleus scattering. 
The production of bubbles is monitored both 
optically and via sound emission, reaching a reconstructed 3D spatial resolution of about 1 mm. This information is crucial 
in rejecting events coming from the walls of the detector, as well as in identifying neutron multiple-scatter events \cite{coupp}. 
For this first phase detector, 
no particular attention was payed to minimize alpha-recoil backgrounds, 
which can for instance be induced by radon and its progeny decays. A total exposure of 250 kg days has been reached, and, in spite of the fact 
that the background was dominated by Rn decays, new, improved limits on spin-dependent WIMP interactions at WIMP masses below 30 GeV/c$^2$ could 
be achieved \cite{coupp_taup07}. New larger chambers, with a much improved radon background, are currently under construction. 
The goal is to operate these at larger depths, at the Soudan Underground Laboratory in Northern Minnesota.
Figure \ref{fig8:coupp} shows the 2\,kg prototype under operation, with single, and multiple bubbles, as well as the projection 
of events in the y-z-plane.

\noindent
{\bf PICASSO and SIMPLE:} the PICASSO experiment is installed at the Sudbury Neutrino Observatory at 2070\,mwe. It uses $\sim$10\,$\mu m$ large droplets of 
a superheated fluorocarbon, C$_4$F$_{10}$, suspended in a polymerized water based gel, as the active WIMP target. The droplets act like 
small bubble chambers, undergoing a phase transition when a certain amount of energy is deposited within a critical length. 
The phase transition depends on the temperature and ambient pressure of the superheated liquid and, most importantly, on the 
specific energy loss of the interacting particle. This allows to suppress the background from particles producing low ionization densities, 
by maintaining full efficiency for strongly ionizing nuclear recoils. Since the energy threshold for nuclear recoils depends on temperature and 
pressure (with lower thresholds for higher temperatures), an energy range of nuclear recoils can be covered by varying the temperature of the 
detector. A phase transition (bubble burst) is read out with piezoelectric transducers mounted on the exterior of the detector wall 
\cite{picasso}. 
The most recent results were obtained from data taken between April-October 2004 with three 1.5\,l detector modules with a total active mass 
of $^{19}$F of $\sim$19.4\,g.
The temperature was varied from 20-47\,C, covering an energy range of $^{19}$F recoils from 6 to 500\,keV, and the effective exposure 
amounted to 1.98\,kg\,days after cuts \cite{picasso}. The temperature dependence of the energy threshold has been studied with neutron, gamma and 
alpha sources, and compared to detailed Monte Carlo simulations of the detector response to these sources. The integrated background over the entire 
recoils spectrum above the 6\,keV threshold of the cleanest detector was $\sim$480\,events\,kg$^{-1}$\,day$^{-1}$, the data being well described 
by the temperature profile expected for alpha particles. Since the expected WIMP response curve differs significantly from the one of the alpha background, 
an upper bound on the WIMP cross section on $^{19}$F was obtained by fitting both contributions simultaneously to the data. The maximum WIMP 
sensitivity is obtained for spin-dependent, pure p-couplings, namely 1.3\,pb at a WIMP mass of 29\,GeV/c$^2$ \cite{picasso}. 
A larger version of the experiment, with improved backgrounds and a total active mass of 3\,kg (32 detectors), is currently being installed 
at SNOLab, with first modules being already in operation. At the current level of intrinsic alpha background, a sensitivity of 
5$\times$10$^{-38}$cm$^2$ for pure-proton, spin-dependent couplings is expected for an exposure of 280 kg days \cite{picasso}. 
SIMPLE is following a similar approach, and has published results from operating four superheated droplet detectors (0.42 kg days exposure) in 2005 
\cite{simple}. The reported limits were almost identical to those published by PICASSO for 2\,kg\,days exposure, namely at a minimum 
of about 1$\times$10$^{-36}$cm$^2$ for pure-proton, spin-dependent couplings. More recent limits were obtained with a prototype CF$_3$I  detector 
with an exposure of 0.024\,kg\,days \cite{simple}.

\section{Conclusions}

A major program to detect the tiny energy depositions when a galactic WIMP scatters off a nucleus in an ultra-low 
background detector is underway. After decades of technological developments, experiments operating 
deep underground have reached the sensitivities to probe realistic supersymmetric particle physics models. 
The major questions  have shifted from 'how to detect a WIMP' to 'how can we identify its nature in case of a signal'. 
A combination of different detector materials, coupled to using target nuclei with and without spin, will allow to 
determine the particle mass, and in some cases, to distinguish among different underlying theoretical models 
(recent studies can be found in \cite{green,cerdeno,drees}).
Needless to say, additional information from accelerators and from indirect searches  
could allow to determine the local density and perhaps to constrain the sub-structure of the dark matter halo.
While recent best limits on WIMP-nucleon cross sections are derived from kg-size experiments, larger, 
100 kg - 1 ton size detectors are already under construction. These experiments will have a non-negligible chance of 
discovering a heavy dark matter particle and their results will strongly influence the design of next generation 
detectors. 

\section*{Acknowledgments}
This work is supported by the 
Swiss National Foundation SNF Grant No 20-118119 and by the Volkswagen Foundation.
I would like to thank the organizers for the invitation to an exciting and stimulating 
conference.

\end{document}